\documentstyle[aps,prl]{revtex}

\draft
\begin{document}
\input{epsf}
\twocolumn[\hsize\textwidth\columnwidth\hsize\csname@twocolumnfalse\endcsname
\title{
$\pi$-kinks in strongly ac driven sine-Gordon systems
}
\author{Vadim~Zharnitsky$^{1}$, Igor~Mitkov$^{2}$,
and Niels~Gr{\o}nbech-Jensen$^{3}$}
\address{
$^{1}$~Division of Applied Mathematics, Brown University,
Providence, RI 02912\\
$^{2}$~Center for Nonlinear Studies and
Computational Science Methods Group\\
Los Alamos National Laboratory,
Los Alamos, NM 87545\\
$^{3}$~Theoretical Division, Los Alamos National Laboratory,
Los Alamos, NM 87545
}

\date{\today}
\maketitle

\begin{abstract}
We demonstrate that $\pi$-kinks exist in non-parametrically ac driven
sine-Gordon systems if the ac drive is sufficiently fast.
It is found that, at a critical value of the drive amplitude,
there are two stable and two unstable equilibria in the sine-Gordon phase.
The pairwise symmetry of these equilibria implies the existence of a
one-parameter family of $\pi$-kink solutions in the reduced system.
In the dissipative case of the ac driven sine-Gordon systems, corresponding
to Josephson junctions, the velocity is selected by the balance between
the perturbations.
The results are derived from a perturbation analysis and verified by direct
numerical simulations.
\end{abstract}
\pacs{
PACS: 03.40.Kf, 52.35.Mw, 02.30.Jr}
\vskip1pc]
\narrowtext

Soliton bearing systems are very important for our understanding of collective
phenomena in many physical systems in the one dimensional approximation
\cite{General_1d}. Often, such systems are perturbed in one form or another
and sometimes these perturbations are temporally
periodic~\cite{AC,cirillo,rotoli}.
If the perturbations are small, one can approximate the dynamics
through the adiabatic perturbation technique \cite{kaup,scott}, where
the integrability of the unperturbed system is used to assume wave profiles
for analytical perturbation techniques applied to the ``nearly integrable''
system. However, are the perturbations large,
we cannot simply assume the unperturbed wave profiles as representing a good
approximation to the dynamics since the near integrability is lost.

This problem was previously addressed for strong periodic perturbations
of sine-Gordon systems \cite{kivshar2,Jensen_93}, where it was shown that
the direct ac drive can induce a Shapiro type
locked phase \cite{Barone_82,miles}
to which a $2\pi$-kink can localize. Similarly, it has been
demonstrated \cite{levi,mitkov,Kivshar_92}
that a parametrically driven
sine-Gordon system can produce stable $\pi$-kink propagation due to the well
known effective Kapitza potential of a driven pendulum \cite{landau}.

It is thereby well documented that strongly perturbed nonlinear soliton bearing
systems can retain some forms of near integrability in certain windows of the
perturbation parameter space. Previous analyses for the sine-Gordon system
have been performed using the analogies to the single pendulum cases, where
Shapiro phase-locking exists for the direct drive, leading to stable
$2\pi$-kinks in a rescaled sine-Gordon chain.

In this paper we demonstrate
that $\pi$-{\it kinks} can propagate in strongly perturbed, directly driven
sine-Gordon chains if the perturbation parameters are chosen near the values,
leading to zero crossing of the relevant Bessel functions determining the
size of a Shapiro step. The analysis is based on the normal form
technique that relies upon a time scale separation
between the rapidly oscillating driving force and a relatively
slow behavior of the residual field.
This technique was previously applied to the parametrically
forced sine-Gordon equation (SGE) in \cite{levi,mitkov}.
We find for nondissipative ac driven SGE a one-parameter family of
$\pi$-kink solutions moving with any prescribed velocity.
In the case of damped and driven SGE, we show that the velocity
is selected and only one of the $\pi$-kinks survives.
We verified the obtained results by performing numerical simulations
of the ac driven SGE.

First, we consider
the equation of motion for a directly forced pendulum
$\ddot \phi + \sin{\phi} = Mf(\omega t)\,,$
where $\phi$ is the phase of the pendulum,
$f$ is a mean-zero periodic function, $M$ is constant,
$t$ represents a normalized time, and
the normalized frequency $\omega$ is assumed to be sufficiently large compared
to the natural frequency of the pendulum, $\omega_0=1$.
We shift the emphasis to the oscillating reference frame
by the transformation
\begin{eqnarray}
\phi=\theta+M\omega^{-2}F(\omega t),
\label{vadim1}
\end{eqnarray}
where $F$ has zero mean and $F^{\prime \prime}(\tau)=f(\tau)$.
We then obtain the parametrically forced equation
$\ddot \theta + \sin{(\theta+M\omega^{-2}F(\omega t))} = 0\,$,
with the Hamiltonian
\begin{eqnarray}
H = \frac{p^{2}}{2}-A(\omega t)\cos(\theta) + B(\omega t)\sin(\theta),
\label{frstham}
\end{eqnarray}
where $p$ is the momentum canonically conjugate to $\theta$, and 
$A(\omega t) = \cos(M\omega^{-2}F(\omega t)),\;\;
B(\omega t) = \sin(M\omega^{-2}F(\omega t))\,$.
Using $2\pi/\omega$ periodicity of $A, B$, we
denote
$\{A\} = A - \langle A \rangle\,,\,
\{B\} = B - \langle B \rangle\,,$
where $\langle\ldots\rangle \equiv (1/2\pi)\int_0^{2\pi} d\tau\ldots$

Following Refs.~\cite{levi,mitkov},
we apply the normal form technique,  to move mean-zero 
terms to a higher order.
Let the first canonical transformation be defined implicitly as
\begin{eqnarray}
p = p_{1}+ \partial_{\theta}W_1 (\theta,p_{1},t), \;\;\;\;\;
\theta_{1}= \theta+\partial_{p_{1}}W_1 (\theta,p_{1},t).
\label{frsttransf}
\end{eqnarray}
The transformed Hamiltonian takes the form
$H_{1} = H + W_{1t}\,.$
To remove mean-zero
rapidly oscillating terms, we choose
$W_1 = \omega^{-1}\{A\}_{-1}\cos(\theta) -
\omega^{-1}\{B\}_{-1}\sin(\theta)\,,$
where $\{ A \}_{-1}$ is a mean-zero antiderivative of $\{ A \}$.
With this choice of $W_1$ the transformed Hamiltonian takes the form
\begin{eqnarray}
H_1 = \frac{p_1^2}{2} &-& \langle A \rangle \cos \theta_1 +  
\langle B \rangle \sin \theta_1
\nonumber\\
&-& \omega^{-1} p_1
\left (  \{A\}_{-1} \sin \theta_1 +
\{B\}_{-1} \cos \theta_1   \right )
\\
&+& \frac{\omega^{-2}}{2}
\left (  \{A\}_{-1} \sin \theta_1 + \{B\}_{-1} \cos \theta_1 \right )^2 .
\nonumber
\end{eqnarray}
Unlike the original Hamiltonian~(\ref{frstham}), the transformed
Hamiltonian $H_{1}$ contains
terms with small ($\sim\omega^{-2}$) positive
time-dependent coefficients $\{A \}^{2}_{-1}, \{B \}^{2}_{-1}$,
which have non-zero
averages for any nontrivial choice of $M, f$, and therefore,
cannot be removed from the Hamiltonian.

These terms have essential effect on the system
dynamics when the lower order potential energy terms vanish, {\em
i.e.} $\langle A \rangle =\langle B \rangle=0$ (all terms $\sim\omega^{-1}$
always have zero averages).
To remove the explicit time dependence from the Hamiltonian up to
the terms $\sim \omega^{-2}$,
we perform a series of canonical transformations
similar to (\ref{frsttransf}) (see also \cite{mitkov}). To remove the terms
$\sim \omega^{-1}$, we apply the transformation 
\begin{eqnarray}
p_1 = p_{2} \!+\! \partial_{\theta_1}W_2 (\theta_1 ,p_{2},t), \;
\theta_{2} = \theta_1 \!+\! \partial_{p_{2}}W_2 (\theta_1,p_{2},t)
\end{eqnarray}
with  
\begin{eqnarray}
W_2 = \omega^{-2} p_2 ( \{A\}_{-2}\sin(\theta_1 ) - \{B\}_{-2}\cos(\theta_1 )).
\end{eqnarray}
After straightforward calculations we obtain the transformed Hamiltonian
\begin{eqnarray}
H_2 &=& \frac{p_2^2}{2}- \langle A \rangle \cos \theta_2 +  
\langle B \rangle \sin \theta_2
\nonumber\\
&+& \frac{\omega^{-2}}{2}
\left (  
 \langle   \{A\}_{-1}^2 \rangle   \sin^2 \theta_2 + 
\langle \{B\}_{-1}^2 \rangle   \cos^2 \theta_2 \right .
\\
&+& \left. \langle   \{A\}_{-1}\{B\}_{-1}  \rangle \sin 2\theta_2
\right )
+\omega^{-2}R + O(\omega^{-3}),
\nonumber
\end{eqnarray}
where $R$ turns out to be mean-zero $\langle R \rangle = 0$. 
Finally, applying the third transformation 
with $W_3= \omega^{-3} R_{-1}$  and neglecting terms $\sim\omega^{-3}$
we obtain
\begin{eqnarray}
\tilde H (P,\Theta) = \frac{P^{2}}{2} - C\cos(\Theta-\gamma)
-\frac{\omega^{-2}}{2}D\cos(2\Theta-\delta),
\label{new}
\end{eqnarray}
where
\begin{eqnarray}
\left. \begin{array}{l}
\langle A \rangle = C \cos(\gamma),\;\;\;
\langle \{A\}_{-1}^{2} \rangle - \langle\{B\}_{-1}^{2}\rangle =
2D\cos(\delta),\\
\langle B \rangle = C \sin(\gamma),\;\;\;
-\langle \{A\}_{-1} \{B\}_{-1} \rangle = D\sin(\delta),
\label{ABC}
\end{array}
\right. \end{eqnarray}
and $P$ and $\Theta$ are new canonical variables.

For $C \neq 0$ there is only one stable equilibrium $\Theta=\gamma$
and one unstable equilibrium $\Theta=\pi + \gamma\,$, for
large frequencies; this equilibrium corresponds to the usual
phase-locked Shapiro state known from Josephson junctions \cite{Barone_82}.
However, as $C$ passes through $0$, a bifurcation occurs and
(for $C=0$) the system has two stable equilibria given by 
$\Theta = \delta/2,\; \pi + \delta/2$ and two unstable equilibria given by
$\Theta = \pi/2+ \delta/2,\; 3\pi/2 + \delta/2\,.$

Now we turn to the {\em directly forced SGE}
\begin{eqnarray}
\phi_{tt}-\phi_{xx}+\sin{\phi}=Mf(\omega t).
\label{dfsge}
\end{eqnarray}
After applying the transformation (\ref{vadim1}) we obtain
the evolution equation for a new phase $\theta$ on top of
a rapidly oscillating background field
(written in the canonical form)
\begin{eqnarray}
\theta_t = p,\;\;\;\;\;
p_t = \theta_{xx} - \sin{(\theta + M\omega^{-2}F(\omega t))}.
\label{dfsge1}
\end{eqnarray}
Invoking  the canonical transformations
similar to those for the
directly forced pendulum (see also \cite{mitkov}) and using
(\ref{new}), we obtain
for the corresponding Hamiltonian
\begin{eqnarray}
H = \int_{-\infty}^{+\infty}\left[
\frac{\Theta_{x}^{2}}{2}
+ \tilde H(P,\Theta)
+ O(\omega^{-3}) \right]dx,
\label{dfsge2}
\end{eqnarray}
where the error terms $O(\omega^{-3})$ contain the derivatives up to
the second order.
For sufficiently large $\omega$, when
we can neglect these terms, 
the above Hamiltonian corresponds to the double SGE.
After retracing the identical transformation
(\ref{vadim1}), the obtained approximate solutions become $\pi$-kinks
on top of the rapidly oscillating background field.

We now consider the {\em damped and driven SGE}
\begin{eqnarray}
\phi_{tt}-\phi_{xx} + \sin{\phi} = Mf(\omega t) - \alpha \phi_{t} +
\eta \; ,
\label{dfdsge}
\end{eqnarray} 
which is frequently used to describe long Josephson
junctions \cite{scott},
where $\phi$ is the phase difference between the quantum mechanical
wave functions of the two superconductors defining the junction, the normalized
time $t$ is measured relative to the inverse plasma frequency, space $x$ is
normalized to the Josephson penetration depth,
and the nonlinear term represents
tunneling of superconducting Cooper pairs, normalized to the critical current
density. The perturbations on the right hand side of the equation represent,
respectively, a normalized ac driving current, a dissipative term arising from
tunneling of quasiparticles, and a normalized dc bias current.

To obtain an effective equation of the evolution on the slow
time scale, we apply the above canonical transformations.
Since the system is no longer hamiltonian, we work with
equations of motion rather than Hamiltonians.
We start with a homogeneous transformation
to the oscillating reference frame,
$\phi = \theta + G(t)\,,$
analogous to~(\ref{vadim1}),
designed to remove the free oscillatory term.
Substituting this transformation to (\ref{dfdsge})
and choosing the function $G$ so that it solves the equation
$\ddot G + \alpha \dot G = Mf(\omega t)\,$,
we obtain the equations of motion in the canonical form
\begin{eqnarray}
\theta_{t} = p, \;\;\;\;\;\;
p_{t} = \theta_{xx} -\alpha p +\eta - \sin(\theta + G(\omega t )).
\label{eqn2}
\end{eqnarray}
For the particular case of $f(\tau) = \sin{\tau}$, we find
\begin{eqnarray}
G(\tau) = -\frac{\alpha}{\omega}\frac{M}{\alpha^{2}+\omega^{2}}\cos{\tau} -
\frac{M}{\alpha^{2}+\omega^{2}}\sin{\tau}.
\label{gfunc}
\end{eqnarray}
Using the notations
$A = \cos{G(\omega t )}\,,\,
B = \sin{G(\omega t )}\,$
and assuming that $\langle A \rangle =\langle B \rangle = 0$,
we apply the series of transformations, as we did
for the directly forced pendulum.
This moves all mean-zero terms to higher order, leading to
\begin{eqnarray}
&&\Theta_{t} = P + O \left(\omega^{-3} \right )
\label{efeq}\\
&&P_{t} = \Theta_{xx} -\alpha P +\eta
-\omega^{-2}D\sin{(2\Theta -\delta)} + O \left(\omega^{-3} \right ),
\nonumber
\end{eqnarray}
where $D$ and $\delta$ are given by (\ref{ABC}).
Constants $\alpha$ and $\eta$ in (\ref{efeq}) are assumed
to be sufficiently small, so that the corresponding terms
could be considered as a perturbation. Then, in zeroth order
in $\alpha, \eta$, the system (\ref{efeq}) reduces (after
neglecting terms $\sim\omega^{-3}$) to SGE,
which has $\pi$-kink solutions.
Therefore, slightly perturbed $\pi$-kinks are
approximate solutions of
the original equation (\ref{dfdsge})
on top of the rapidly oscillating background field.

To verify the above predictions,
we have performed a set of numerical simulations.
Rescaling the variables in equations of motion~(\ref{efeq}) (which in
Hamiltonian case correspond to the Hamiltonian
(\ref{dfsge2})), by
\begin{eqnarray}
\left. \begin{array}{l}
\chi = 2\Theta-\delta,\;\;\;\;\;\;\;\;
p_{\chi} \frac{\sqrt{D/2}}{\omega} = P, \\
X=\frac{\sqrt{2D}}{\omega}x,\;\;\;\;\;\;\;\;
T=\frac{\sqrt{2D}}{\omega}t,
\label{rescale}
\end{array} 
\right. \end{eqnarray}
we obtain the sine-Gordon system
\begin{eqnarray} 
\chi_{\scriptscriptstyle T} = p_{\chi},\;\;\;\;\;
p_{\chi {\scriptscriptstyle T}} = \chi_{{\scriptscriptstyle XX}} -\alpha
\frac{\omega}{\sqrt{2D}}p_{\chi} + 
\frac{\omega^{2}}{D}\eta - \sin{\chi}.
\label{sinegordon}
\end{eqnarray}

\noindent
In the {\bf Hamiltonian case} ($\alpha = \eta = 0$),
(\ref{sinegordon}) has solitary wave solutions
which, after retracing the transformation~(\ref{rescale}), read
$\Theta= V(x,t),\; P=V_{t}(x,t),$ where
\begin{eqnarray}
V(x,t) = \frac{\delta}{2} + 
2\arctan\left[\exp{\frac{\sqrt{2D}}{\omega}\left(\frac{x-ct}
{\sqrt{1-c^{2}}}\right)}\right].
\label{icv}
\end{eqnarray}
Using (\ref{frsttransf}) we return to the original variables
which gives us the approximate solution
\begin{eqnarray}
\theta = V,\;\;\;
p = V_{t} - \frac{\{A\}_{-1}}{\omega}\sin(V)
- \frac{\{B\}_{-1}}{\omega}\cos(V),
\label{igor1}
\end{eqnarray}
which we use to generate initial conditions
for~(\ref{dfsge1}).

\begin{figure}[h]
\hspace{-2cm}
\rightline{ \epsfxsize = 7.0cm \epsffile{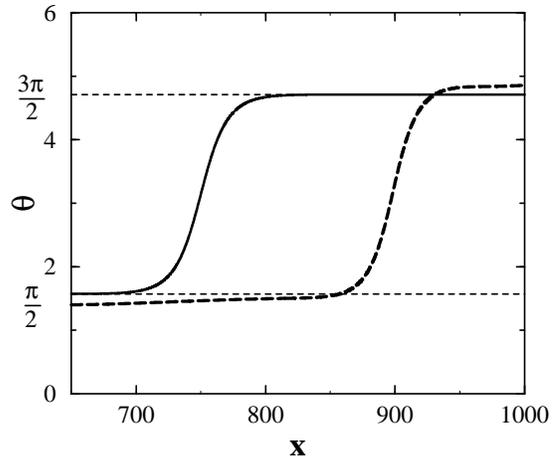}}
\caption{
The behavior of $\pi$-kink solution in the
ac driven SGE (\protect\ref{dfsge1}). The solid line
corresponds to the initial profile given by (\protect\ref{icv}),
with $c = 0.5$ and $F(\omega t) = \protect\sin(\omega t)$.
The dashed line corresponds to $t=300$. The parameters
are $\omega = 10$, time step $dt = 0.01$, mesh size $dx = 0.1$,
system size $L = 1500$.
\label{fig1}}
\end{figure}

We have performed numerical simulations of~(\ref{dfsge1}),
which is equivalent to the original directly forced
SGE (\ref{dfsge}). We used a second-order leap-frog method,
with initial conditions given by (\ref{igor1}).
The coefficients $D$ and $\delta$ in~(\ref{icv})
are calculated from~(\ref{ABC}): $\;D = 0.2270596\ldots\,;\;
\delta \approx \pi\,$.
Figure~\ref{fig1} shows the results of the simulations.
The driving amplitude $M$ was chosen so as to make both
$\langle A \rangle$ and $\langle B \rangle$ vanish:
$M\omega^{-2}\approx 2.4048\ldots$. The velocity $c$ in~(\ref{icv})
was taken as $c = 0.5$. The small parameter
used in our perturbation analysis is $\epsilon = \omega^{-1}$. 
One can see from the Figure that the kink indeed moves with
the velocity $c\approx 0.5$ and that it is stable
for $t \leq 300\,$, much longer than $\epsilon^{-1}$
(in the considered case $\epsilon=0.1$),
which is a natural estimate for the validity of
averaging for time-periodic perturbation
based on multiple scale procedure. For longer times
($t \geq 300$), the $\pi$-kink becomes unstable
and eventually disintegrates due to the parametric resonances. 
A similar destruction of the $\pi$-kink after a long time
was observed in a parametrically excited
SGE~\cite{levi,mitkov}.

\vspace{1.0cm}

\vspace{0.3cm}
\noindent
In the {\bf dissipative case} ($\alpha \neq 0$ and $\eta \neq 0$),
we have shown that equation (\ref{dfdsge}) reduces to (\ref{efeq}),
after averaging over the fast time scale and neglecting terms
of order $\omega^{-3}$.
Then only one $\pi$-kink solution of (\ref{efeq}),
with a certain value of $c$,
is selected out of the entire family,
because of the energy balance consideration.

To find the selected $c$, we substitute a traveling wave ansatz,
$Z = X - c T$, into (\ref{sinegordon}).
In the zeroth order in $\alpha$ and $\eta$, we obtain
$\chi_{\scriptscriptstyle 0} =
4 \arctan \left[\exp\,(Z/\sqrt{1-c^2})\right].$
The solvability condition for the linearized equation
for the first order correction $\chi_{\scriptscriptstyle 1}$,
gives us the velocity
\begin{eqnarray}
c = -\,\frac{\sqrt{2}\omega\eta}{\alpha\sqrt{D}}\;
\frac{\int_{-\infty}^{+\infty} \chi_{\scriptscriptstyle 0}^{\prime} dZ}
{\int_{-\infty}^{+\infty} (\chi_{\scriptscriptstyle 0}^{\prime})^2 dZ}
= -\,\frac{\pi\eta\omega}{\sqrt{8\alpha^2 D + \pi^2\eta^2\omega^2}}\;.
\label{igor7}
\end{eqnarray}

\begin{figure}[h]
\hspace{-1.5cm}
\rightline{ \epsfxsize = 7.0cm \epsffile{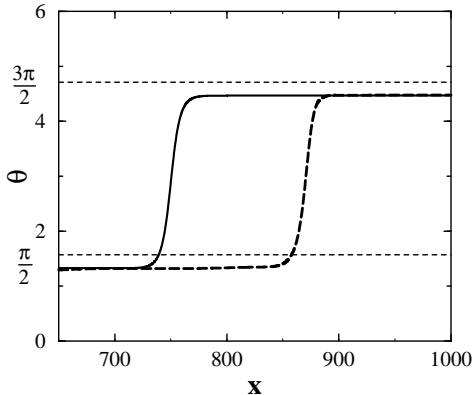}}
\caption{
The behavior of $\pi$-kink solution in the
damped and driven SGE (\protect\ref{eqn2}). The solid line
corresponds to the initial profile
$\theta(x,0) = V(x,0) + \protect\arcsin(\omega^2\eta/D)$,
with $V$ given by (\protect\ref{icv})
and $G(\omega t)$ given by (\protect\ref{gfunc}).
The dashed line corresponds to $t=150$. The parameters
are $\omega = 6$, time step $dt = 0.01$, mesh size $dx = 0.1$,
system size $L = 1500$, $\alpha = 0.03$, $\eta = -0.003$.
\label{fig2}}
\end{figure}

We have simulated (\ref{eqn2}), equivalent to the original 
forced and damped SGE~(\ref{dfdsge}), with the initial 
conditions obtained from (\ref{igor1})
by adding a small correction $\arcsin(\omega^2\eta/D)$
to the first equation of (\ref{igor1}), to compensate
an additional energy transferred to the kink from the constant
source in (\ref{eqn2}).
The results of the simulations are given in Figure~\ref{fig2}.
The $\pi$-kink in the Figure moves with the velocity
$c \approx 0.79\,,$ which approximately coincides with the
selected velocity $c \approx 0.81$ obtained from~(\ref{igor7}).
The plotted solution remained intact for times $t \le 150$.
Shortly after that it is destroyed by higher order effects such as radiation
and nonlinear resonances.

We have demonstrated that directly strongly ac driven
SGE can produce localized $\pi$-kinks in special
regions of parameter space. These regions coincide with the regions where
$2\pi$-kink localization (Shapiro steps) vanishes. The formalism for
demonstrating the existence of the $\pi$-kinks is based on a time scale
separation technique, where the ac drive is assumed to be fast compared to
any natural oscillation in the unperturbed SGE. As a consequence,
the predicted localization mechanism is limited to the high frequency region
of parameter space. As the driving frequency is lowered, nonlinear mixing
between the supposed high frequency drive and low frequency wave dynamics
become more dominant, eventually leading to the destruction of localization
and all coherent behavior. We have numerically tested that long
time propagation
of $\pi$-kinks is possible for relatively low driving frequencies,
as small as $\omega\sim 5$, and moderate driving amplitudes, $M\sim 60$.
While technical
Josephson applications of $\pi$-kink localization seem far removed from the
present, the above perturbation parameters suggest that the predicted effect
may exist within experimental parameters used for generating Shapiro steps
in current-voltage characteristics of Josephson junctions.

This work was performed under the auspices of the U.S.\ Department of Energy.
Work of VZ was supported by NSF under Grant No. DMS-9627721.


\begin{thebibliography}{99}

\bibitem{General_1d} see, e.g., Yu.\ S.\ Kivshar and B.\ A.\ Malomed,
Rev.\ Mod.\ Phys.\ {\bf 61}, 763 (1989).

\bibitem{AC}M.\ Salerno, M.\ R.\ Samuelsen, G.\ Filatrella, S.\ Pagano,
and R.\ D.\ Parmentier, Phys.\ Rev. {\bf B 41}, 6641 (1990).

\bibitem{cirillo} M.\ Cirillo and F.\ L.\ Lloyd,
J.\ Appl.\ Phys.\ {\bf 61}, 2581 (1987).

\bibitem{rotoli} G.\ Rotoli, G.\ Costabile, and
R.\ D.\ Parmentier, Phys.\ Rev. {\bf B 41}, 1958 (1990).

\bibitem{kaup} D.\ J.\ Kaup and A.\ C.\ Newell,
Phys.\ Rev. {\bf B 18}, 5162 (1978).

\bibitem{scott} D. W. McLaughlin and A. C. Scott,
Physical Review {\bf A 18}, 1652 (1978).

\bibitem{kivshar2} Niels Gronbech-Jensen and Yuri S. Kivshar, 
Phys.\ Lett. {\bf A 171}, 338 (1992).

\bibitem{Jensen_93} Niels Gronbech-Jensen, Yuri S. Kivshar, Mario Salerno,
Phys.\ Rev.\ Lett. {\bf 70}, 3181 (1993).

\bibitem{Barone_82} see, e.g., A.\ Barone and G.\ Patern\'{o},
{\it Physics and applications of the Josephson effect} (Wiley, New York, 1982).

\bibitem{miles} J. Miles, Phys. Lett. {\bf A 133}, 295(1988).

\bibitem{levi} V. Zharnitsky, I. Mitkov, M. Levi,
Phys. Rev. {\bf B 57}, 5033 (1998).

\bibitem{mitkov} I. Mitkov, V. Zharnitsky, to appear in Physica
{\bf D} (1998).

\bibitem{Kivshar_92} Yu.\ S.\ Kivshar, N.\ Gr{\o}nbech-Jensen, and
M.\ R.\ Samuelsen,
Phys.\ Rev. {\bf B 45}, 7789 (1992).

\bibitem{landau} see, e.g., L.D. Landau, M. Lifshitz,
Mechanics, Pergamon Press, 
Oxford 1960,

\end{thebibliography}
\end{document}